# Magnetotransport in graphene/Pb$_{0.24}$Sn$_{0.76}$Te heterostructures: finding a way to avoid catastrophe

Gregory M. Stephen[1*], Ivan Naumov[2], Nicholas A. Blumenschein[1], Y.-J. Leo Sun[3], Jennifer E. DeMell[1], Sharmila N. Shirodkar[2], Pratibha Dev[2], Patrick J. Taylor[4], Jeremy T. Robinson[5], Paul M. Campbell[5], Aubrey T. Hanbicki[1], Adam L. Friedman[1*]

[1]Laboratory for Physical Sciences, 8050 Greenmead Dr. College Park, Maryland 20740, USA;
[2]Howard University, Department of Physics and Astronomy, Washington, D.C. 20059, USA;
[3]University of Maryland, Department of Electrical and Computer Engineering, College Park, Maryland 20742, USA;
[4] US Army Research Laboratory, 2800 Powder Mill Rd., Adelphi, MD 20783, USA;
[5] US Naval Research Laboratory, Electronics Science and Technology Division, 4555 Overlook Ave., S.W., Washington, DC 20375, USA;

[*]Corresponding Authors: Gregory M. Stephen *gstephen@lps.umd.edu*, Adam L. Friedman *afriedman@lps.umd.edu*

## Abstract

While heterostructures are ubiquitous tools enabling new physics and device functionalities, the palette of available materials has never been richer. Combinations of two emerging material classes, two-dimensional materials and topological materials, are particularly promising because of the wide range of possible permutations that are easily accessible. Individually, both graphene and Pb$_{0.24}$Sn$_{0.76}$Te (PST) are widely investigated for spintronic 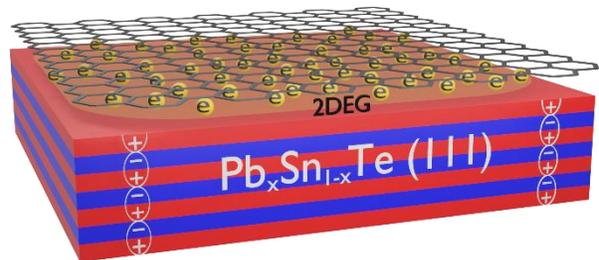 applications because graphene's high carrier mobility and PST's topologically protected surface states are attractive platforms for spin transport. Here, we combine monolayer graphene with PST and demonstrate a hybrid system with properties enhanced relative to the constituent parts. Using magnetotransport measurements, we find carrier mobilities up to 20,000 cm$^2$/Vs and a magnetoresistance approaching 100%, greater than either material prior to stacking. We also establish that there are two distinct transport channels and determine a lower bound on the spin relaxation time of 4.5 ps. The results can be explained using the polar catastrophe model, whereby a high mobility interface state results from a reconfiguration of charge due to a polar/non-polar interface interaction. Our results suggest that proximity induced interface states with hybrid properties can be added to the still growing list of behaviors in these materials.

Heterostructures are fertile hunting ground for novel properties that are useful for applications or solutions to a variety of physics and engineering problems. Examples include high-mobility two-dimensional electron gases (2DEGs) at the interface between complex oxides caused by charge redistribution to avoid the polar catastrophe,[1–4] the development of blue LEDs in InGaN quantum wells,[5] and the fractional quantum Hall effect observed in two-dimensional (2D) electron systems,[6] the latter two of which resulted in Nobel prizes. In heterostructures such as these, hybridization and proximity effects often dominate interactions yielding unforeseen and extraordinary results. Until recently, creating such heterostructures required careful, spotlessly clean growth and interfaces through molecular beam epitaxy (MBE) or other methods.[7] These techniques rely on understanding free energies of formation and are compatible only with carefully paired materials. However, with 2D van der Waals materials like graphene,[8] and topological materials,[9] these legacy rules no longer apply. Exfoliatable 2D materials can be positioned and stamped arbitrarily on any substrate. Meanwhile, topological materials have spin-orbit coupling (SOC) that is strong enough to create topologically protected and robust surface states that allow for proximity interactions even without pristine interfaces.[10]

Future computing devices provide the most compelling motivation for devising new heterostructures. As current computing technologies approach the inevitable physical and energetic limitations imposed by CMOS-based devices, new approaches to memory and logic implementation are paramount.[11] Alternate state variables such as the electron spin and quantum phase are gaining traction as avenues for memory and logic applications.[12–14] It is difficult to find individual materials with all the properties necessary for alternative state variable computing applications. However, we can design composite systems with the required properties by combining materials into heterostructures. For example, because spin diffusion and relaxation are directly related to the carrier mobility, high-mobility materials such as graphene are obvious candidates for useful spintronic devices.[15,16] While graphene has the long spin diffusion length and relaxation times necessary for spin transport, it lacks the SOC needed to manipulate spin through, *e.g.*, external gating that exploits the Rashba effect.[17] On the other hand, topological insulators (TIs) possess the SOC essential for controlling spin transport as well as topologically protected surface states that reduce the influence of surface defects.[9] Indeed, previous research shows that graphene/TI heterostructures exhibit proximity-induced SOC effects using BiSe-based TIs.[10,18] However, TIs are often heavily doped away from the Dirac point. This introduces significant bulk conduction, leading to mobilities that seldom exceed 1,000 cm$^2$/Vs.[19,20] Combining graphene and topological materials could yield a heterostructure that exhibits the protected, spin-polarized states of the TI with the high mobility of graphene.

Along this line, we fabricate heterostructures of the topological crystalline insulator (TCI) Pb$_{0.24}$Sn$_{0.76}$Te (PST) and graphene (Gr) and find evidence of a high-mobility 2D interface conduction state. See the Supplemental Information for detailed methods. The strong interaction between the component materials leads to enhanced properties such as a mobility and magnetoresistance greater than the constituent materials. We attribute these enhancements to conduction through a 2DEG formed at the interface. Density functional theory (DFT) analysis reveals that the 2DEG does not inherit the topological character of the PST, which is actually broken by adding graphene to the surface. Rather, the 2DEG arises from a charge redistribution at the interface to avoid the polar catastrophe at the surface, similar to the widely studied 2DEG in LaAlO$_3$/SrTiO$_3$.[21] In addition to electrical characterization, in this work we characterize this hybrid system with both high and low field magnetotransport measurements and estimate spin-based properties, such as lifetimes based on both Shubnikov-de Haas and localization effects. Engineered quantum materials such as this could provide a platform for next-generation spintronic devices.

**Results/Discussion**

Fig. 1(a) is an optical image of a measured Hall bar device with a schematic of the heterostructure in the inset. Fig. 1(b) shows the Raman *G* and *2D* peak positions vs. widths for Gr/PST (blue circles) as well as a witness Gr/SiOx (red squares) sample with graphene prepared in the same way for both samples. In this figure, the small symbols are data taken at multiple locations over extended regions of the samples and the large symbols represent averages of these data points with the error bars representative of the statistical mean of the data. For the Gr/PST, both the *G* and *2D* peaks shift to lower energy and the *G* broadens. This is indicative of a reduction of carriers in the graphene,[22] and is consistent with carriers transferring to a surface state between the graphene and PST. Additionally, there is no evidence of strain in the Gr/PST system that could affect the band structure. Strain manifests as a splitting of the *G* peak as well as a correlation of the *2D* and *G* peak positions.[23] We do not observe any splitting of the *G* peak (Fig. 1(b)-inset) and a fit to the *2D* peak position vs. *G* peak position is not consistent with strain (see supplemental Fig. S1(b)).

This notion of a combined system is supported by DFT calculations of the charge distribution. We consider two extreme cases: (i) a heterostructure of graphene on the purely topological SnTe and (ii) a similar heterostructures with the top four layers of Sn replaced with Pb—the latter case is equivalent to having four layers of the trivial insulator PbTe on top of the TCI SnTe and is roughly equal to the stoichiometry of our PST film (See Supplemental Information Fig. S2). This represents a fully phase-segregated alloy of $Pb_{0.24}Sn_{0.76}Se$ where the TCI phase is completely broken at the interface by trivial PbTe. Fig. 1(c,d) shows changes in the electron distribution for graphene on (c) SnTe, and (d) 4L-PbSnTe along with the position of each atomic species. Adding graphene breaks band inversion, creating the charge redistribution observed in Fig. 1(c). By adding enough Pb, the SnTe band inversion is broken and there is no resulting charge density wave created by the graphene, Fig. 1(d).

While the broken band inversion destroys the topological surface states, an alternate surface state is created by the PST surface. SnTe is a mirror TCI with a mirror Chern number $(C_{+i} - C_{-i})/2=2$.[24] Since the total Chern number $(C_{+i} + C_{-i})$ is zero, the Wannier functions in this system can be chosen to be maximally localized in all three dimensions. Therefore, the electrical polarization, which is expressed in terms of the Berry phase, is a well-defined quantity.[25] The calculated total (electronic plus ionic) Berry phase along the [111] direction in the SnTe system is $\pi$, indicating that the two (top and bottom) (111) surfaces are polar.[2,3,26] These surfaces remain polar even if some of the Sn ions are replaced with Pb ions. For the polar surface to be electrostatically stable, it must have a finite external compensating charge density $\sigma_{ext}$, such that $\sigma_{ext} + \mathbf{P} \cdot \mathbf{n} = 0$, where $\mathbf{P}$ is the bulk polarization proportional to the total Berry phase.[2] In the case of the SnTe slabs with Te-termination on both sides, the structure is inversion-symmetric, although it has a broken bulk stoichiometry. For such a Te-terminated slab, $\sigma_{ext}$ comes from the partial filling of the surface states that fall out of the bulk valence band. Because of the presence of two surfaces related by an inversion center, these surface states localized on the opposite surfaces are doubly degenerate. In principle, they can be removed from the energy window around the Fermi level by passivation (for example, with H-atoms) of both surfaces. However, when one of the two (111) Te-terminated surfaces is in contact with graphene, the corresponding surface states (from graphene side) survive, although modified. Our DFT analysis demonstrates that the metallic surface states under the graphene layer are not confined to the very top layer but extend into the bulk region over ~10 layers (see Supplemental Information Figs. S3 and S4). Thus, a 2DEG is created, similar to the LaAlO$_3$/SrTiO$_3$ interface.[2,3] This 2DEG appears in both Fig. 1(c) and (d), indicating that it is not a direct result of the topological state. Rather, bringing graphene into contact with a PST film modifies not only the surface states associated with $\sigma_{ext}$ but also leads to the redistribution of the electron density across the entire slab.

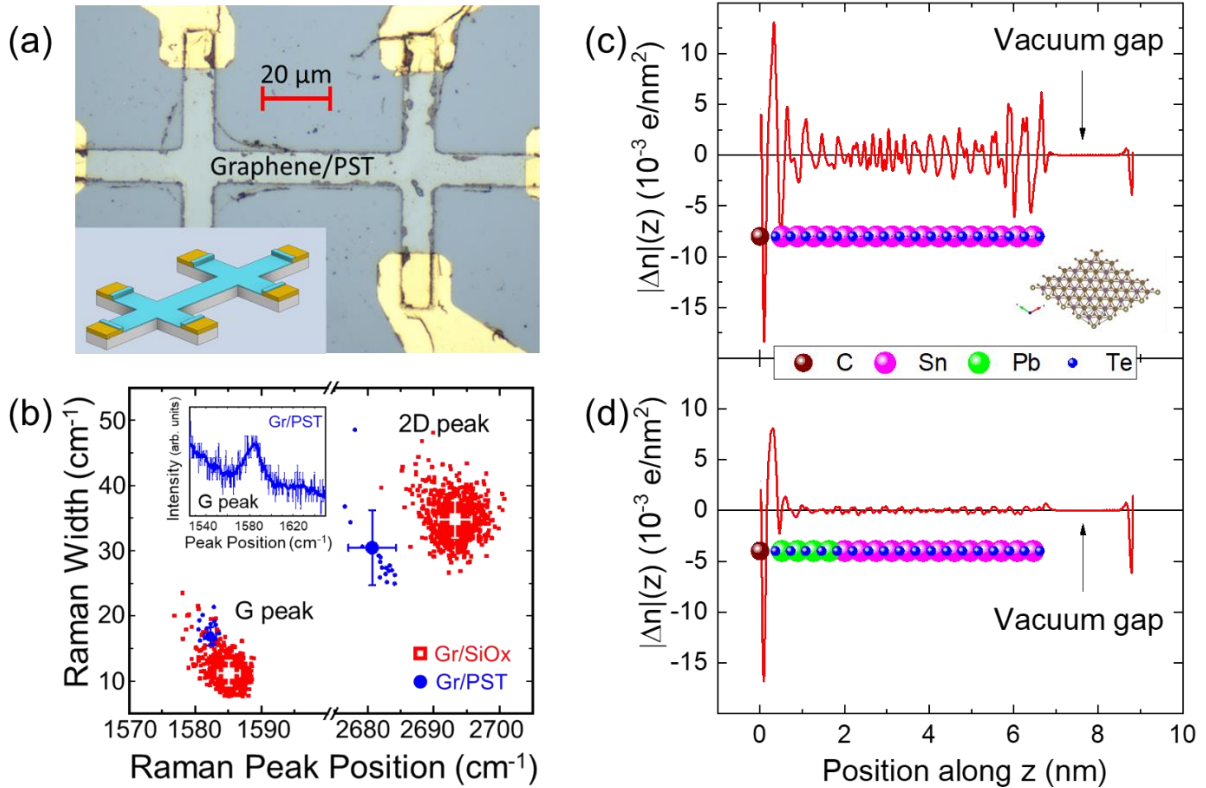

FIG. 1. (a) Optical image of the Gr/PST Hall bar. Inset is a schematic of the heterostructure. (b) Graphene *G* and *2D* Raman peak positions vs. widths for Gr/SiOx (red squares) and Gr/PST (blue circles). The small symbols are data taken over an extended region of the sample and the large data point is the average of these points. The error bars represent the mean of the averaged data. The inset is a Raman spectrum around the *G* peak for the Gr/PST sample. For a strained sample, the *G* peak should exhibit a splitting.[23] (c-d) Change in planar charge density with the addition of graphene on (c) SnTe and (d) SnTe with the top four layers of Sn replaced by Pb. The circles represent the position of each atom in the supercell. The monolayer graphene induces a charge density wave in the SnTe stack, and the addition of Pb layers drastically attenuates the effect of the graphene.

To understand the electronic properties of this heterostructure, we measured the magnetotransport of graphene, PST, and the Gr/PST heterostructure using standard Hall bar devices. To begin, sheet resistances are 36 kΩ/□ for the PST and 4.2 kΩ/□ for the graphene, whereas the heterostructure has a sheet resistance of 3 kΩ/□. This suggests conduction is not just a result of parallel conduction channels which would yield a sheet resistance of 3.76 kΩ/□. The Hall mobility at 10 K for our bare PST film is 200 cm$^2$/Vs and for our CVD graphene is 3,800 cm$^2$/Vs. However, the heterostructure mobility is measured as high as 20,000 cm$^2$/Vs. The quality of the substrate affects the mobility of graphene, with atomically smooth hexagonal boron nitride (hBN) being a common method of improving device quality.[27] However, the physics that allows for increased mobility on hBN (atomically smooth, uniform insulator) do not apply to the doped PST film used here. The sheet carrier densities for the graphene and Gr/PST are within a factor of two (graphene = $4 \times 10^{12}$ cm$^{-2}$; PST = $8 \times 10^{12}$ cm$^{-2}$) suggesting that carrier doping is insufficient to account for the drastic change in mobility. As discussed above, a more likely explanation is a highly-conductive channel created through the avoidance of the polar catastrophe. As shown, the interaction between the two materials affects the charge density throughout the entire sample through electronic relaxation, not just at the surface or interface, although the conductive channel is formed at the surface due to the accumulation of charge in that region.

Magnetoresistance (MR) measurements reveal further insight into the character of the conduction in the surface channel. High field MR of graphene (red), PST (black) and Gr/PST (blue) at T = 300 mK are shown in Fig. 2(a). The heterostructure shows drastically different behavior compared to the constituent materials. We first note that the PST MR (black line) saturates at 6% with no evidence of Shubnikov-de Haas (SdH) oscillations while the graphene shows linear MR with sinusoidal oscillations starting around 7 T, and a similar overall magnitude as the bare PST. The heterostructure exhibits a combination of the two behaviors, but with drastically higher MR approaching 100%. The MR is nearly linear up to 2 T, at which point low-frequency SdH oscillations begin. The background at higher fields appears to be sub-linear, similar to the PST-only sample. The linear behavior at lower fields could be an indication of topological features as seen in Bi$_2$Te$_x$Se$_{1-x}$ topological insulators.[28,29] Quantum effects like SdH oscillations tend to appear when the quantity $\omega_c \tau > 1$, where $\omega_c$ is the cyclotron frequency and $\tau$ is a characteristic scattering time. Using the Drude transport model, this can be written as $\mu B > 1$, where $\mu$ is the carrier mobility and $B$ is the applied field.[30] In other words, oscillations occur when the carriers complete, or nearly complete, cyclotron orbits between momentum scattering events. Thus, the SdH oscillations, or lack thereof, in each sample is well described by the Hall mobility of each sample. Again, the Hall mobility at 10 K for our bare PST film is 200 cm$^2$/Vs, compared to 3,800 cm$^2$/Vs for the graphene and as high as 20,000 cm$^2$/Vs for the heterostructure. From the Drude model, this results in the expected onset of SdH oscillations at fields of approximately 0.5 T and 7 T for Gr/PST and graphene, respectively, and 50 T for bare PST. This is consistent with the data, where oscillations begin in the heterostructure at about 2 T, the graphene around 7 T, and are not observed in the PST. Quantum mobility can also be extracted directly from the SdH oscillations. The quantum mobility is related to the slope of Landau level position vs. number and yields 970 cm$^2$/Vs for the graphene and 8,040 cm$^2$/Vs for the heterostructure, consistent with the low-field Hall mobilities.

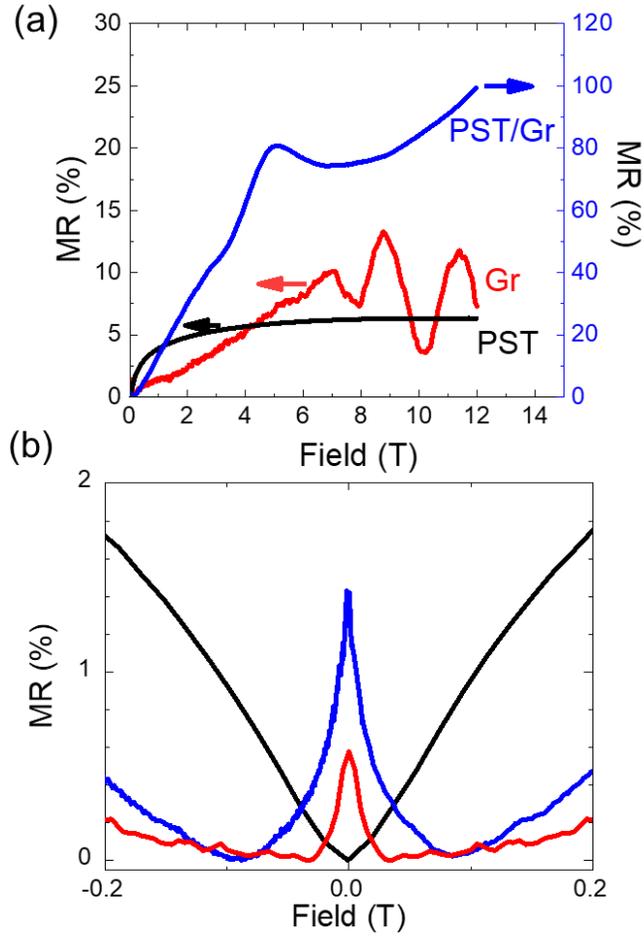

FIG. 2. (a) Magnetoresistance (MR) of PST (black), graphene (red) and the Gr/PST heterostructure (blue). SdH oscillations do not appear in the PST, but begin around 7 T in the graphene and 2 T in the heterostructure. This is explained in part by the large difference in carrier mobilities between the three samples. Only positive field data is shown to emphasize the oscillations. Features are symmetric in field. (see Supplemental Information Figs. S6, S7) (b) A magnified view of the low-field region showing the weak localization (WL) peak in the Gr/PST and graphene, as well as the lack thereof in PST.

Fig. 2 shows the magnetoresistance for both high and low magnetic fields, demonstrating SdH oscillations in Fig. 2(a) and weak localization in Fig. 2(b). At higher fields, SdH oscillations appear on top of the MR signal. These oscillations result from field splitting of the Landau levels and are governed by the momentum scattering in the system. To isolate the oscillatory component, we subtract a background signal from each longitudinal resistance measurement (see Supplemental Information Fig. S3). Fig. 3(a) shows the results of this background subtraction, $\Delta R_{xx}$, as a function of $1/B_z = 1/B\sin\theta$ for angles from 10° to 90°. The angle corresponds to orientation of the B field with respect to the plane of the PST film. Note the peak positions are constant as a function of $1/B_z$. This indicates coupling only to the z-component of the magnetic field and implies conduction is confined to a 2D state, again consistent with a graphene/PST interface state.

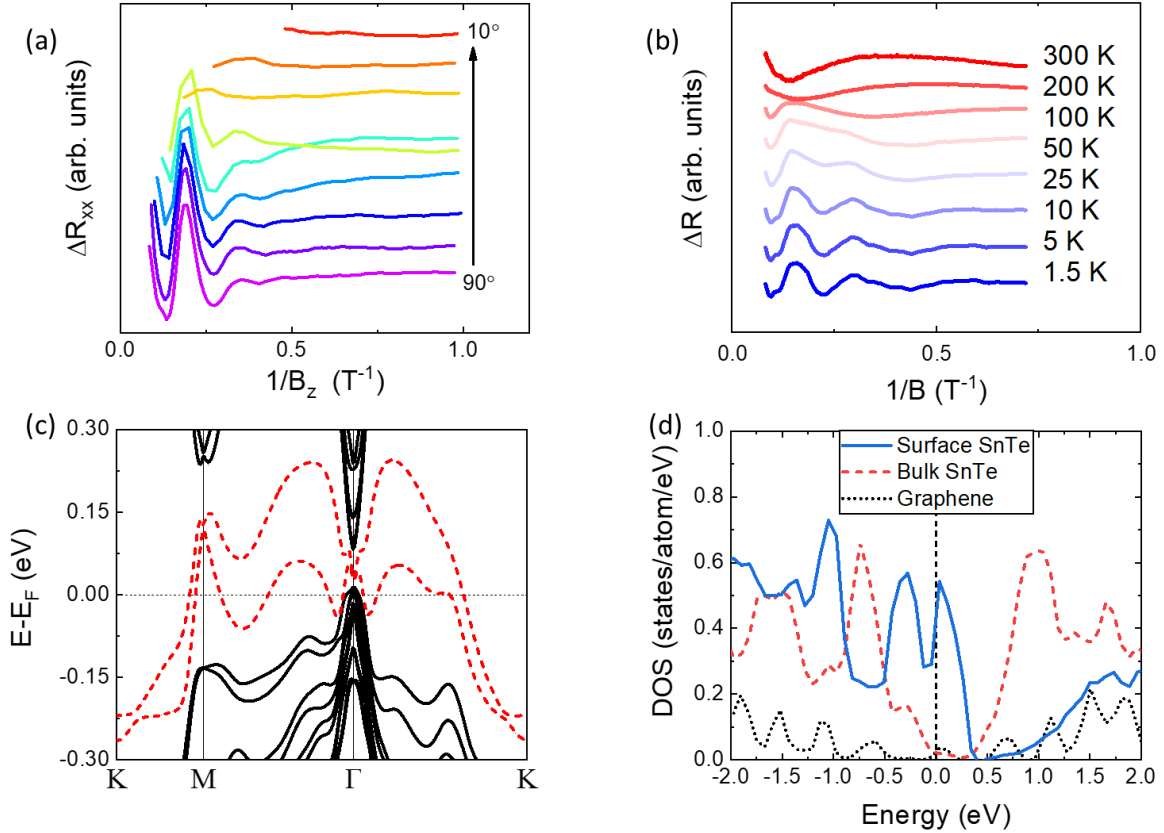

FIG. 3. (a) Background subtracted longitudinal resistance measurement, $\Delta R_{xx}$, as a function of $1/B\sin\theta = 1/B_z$ in Gr/PST for various angles at 300 mK. The SdH peak positions are constant as a function of $1/B_z$, indicating we are probing a purely 2D state. (b) Background subtracted longitudinal resistance measurement, $\Delta R_{xx}$, as a function of $1/B$ in Gr/PST for various temperatures at a 90° angle (out-of-plane field). The peak positions remain constant in temperature, but with decreasing amplitude. Fitting of the field dependence at 1.5 K results in a momentum scattering time of 46 fs, with temperature dependent fitting yielding an effective mass of $0.0095m_e$. These are 50x and 5x larger than the values for graphene, 1.4 fs and $0.002m_e$, respectively. (c) Calculated band structures for Gr/SnTe. Dashed lines represent the bands corresponding to the surface 2DEG. (d) Partial density of states (DOS) for the graphene (dotted black) as well as the SnTe surface (solid blue) and bulk (dashed red). The majority of available states at the Fermi level come from the surface.

Fig. 3(b) shows the background-subtracted data at temperatures from 1.5 K up to 300 K. The temperature dependence of the SdH amplitude, $A$, is directly related to the effective carrier mass, $m^*$. This temperature dependence at field $B_r$ is described by[31]

$$A \propto \lambda T / \sinh \lambda T \qquad (1)$$

where $\lambda = 2\pi^2 k_B m^* / \hbar e B_r$. Fits of our data yield an effective mass of $m^* = 0.0095 m_e$, similar to effective masses previously reported in graphene/WS$_2$ stacks (see Supplemental Information Fig S8).[32] When applied to the oscillations in our graphene sample, we see an effective mass of $m^* = 0.002 m_e$. A 5x increase in effective mass in the heterostructure seems at odds with the increased Hall mobility in the heterostructure.

This discrepancy in effective mass can be reconciled by examining the scattering time. At a fixed temperature, the decay of the oscillations with 1/B is related to the momentum scattering time $\tau_q = \tau_{SdH}$ according to[31]

$$\ln\left(AB \sinh \frac{2\pi k_B T m^*}{eB\hbar}\right) = C - \frac{\pi m^*}{e\tau_{SdH}} \frac{1}{B} \qquad (2)$$

where *A* is the amplitude of a resonance at field *B* and *C* is a constant. With the limited number of resonances observed, the measured value should be considered a rough estimate. From the plot of Eq. 2, the Dingle plot, we extrapolate a scattering time of 46 fs at 1.5 K, much longer than our graphene value of 1.4 fs. Together, the 50x longer scattering time combined with the 5x heavier effective mass is consistent with the order of magnitude increase in the mobility with the addition of PST.

Fig. 3(c) shows the DFT band structure for Gr/SnTe (111) interface. The graphene Dirac cone at Γ is gapped due to hybridization with the SnTe. The highlighted bands (dashed lines) correspond to the 2DEG and originate from the top 2 layers of Sn and Te. Interestingly, these bandstructure calculations resemble the case of unpassivated SnTe,[33] with the important difference being the graphene breaks the mirror symmetry resulting in gaps of the Dirac cones and a loss of the topological state. The transport is thus dominated by the 2DEG that crosses the Fermi energy and is a result of the electronic redistribution to avoid the polar catastrophe. These band structure calculations show that conduction does not happen solely through the graphene layers, but rather through this high mobility channel, consistent with experimental observations. This is even clearer when looking at the densities of states (DOS) shown in Fig. 3(d). The SnTe surface (solid blue line) dominate the DOS around the Fermi level, with each atom contributing nearly 20x more states than either the graphene (dotted black line) or bulk SnTe (dashed red line).

At low field, a cusp can appear in the resistivity that is governed by the scattering mechanisms within the conduction channel. This could be weak localization (WL), commonly seen in graphene, or weak antilocalization (WAL), common in topological materials, the difference being whether the electrons interfere constructively or destructively. While there is no clear localization peak in the PST sample, suggesting the film is heavily doped into the conduction band, WL appears in both graphene and Gr/PST samples, as seen in Fig. 2(b). As we expect conduction to occur through the graphene due a conductivity that is orders of magnitude higher than the PST, we model the cusp using the 2D WL model most often employed to analyze graphene, which accounts for spin-orbit, intervalley and phase coherence scattering:[34–36]

$$\Delta\sigma_{xx}(B) = -\frac{e^2}{2\pi h}\left[F\left(\frac{\tau_B^{-1}}{\tau_\phi^{-1}}\right) - F\left(\frac{\tau_B^{-1}}{\tau_\phi^{-1}+2\tau_i^{-1}}\right) - 2F\left(\frac{\tau_B^{-1}}{\tau_\phi^{-1}+\tau_{so}^{-1}}\right)\right] \quad (3)$$

Where $F(x) = \ln(x) + \psi(1/2 + x^{-1})$, and with $\psi(x)$ being the digamma function. The parameter $\tau_B^{-1} = \frac{4qD}{\hbar}B$, $\tau_\phi$ is the phase-coherence time, $\tau_i$ the intervalley scattering time, and $\tau_{SO}$ is the spin-orbit scattering time. As is typical, the WL peak decreases as a function of temperature, with the feature barely visible at 50 K and absent in the 100 K data, as shown in Fig. 4(a). We can combine these three scattering mechanisms to get a characteristic low field scattering time $\tau_{WL}^{-1} = \tau_i^{-1} + \tau_\phi^{-1} + \tau_{SO}^{-1}$, as shown in Fig. 4(b) (red line). A temperature dependence of the constituent scattering times is shown in the inset.

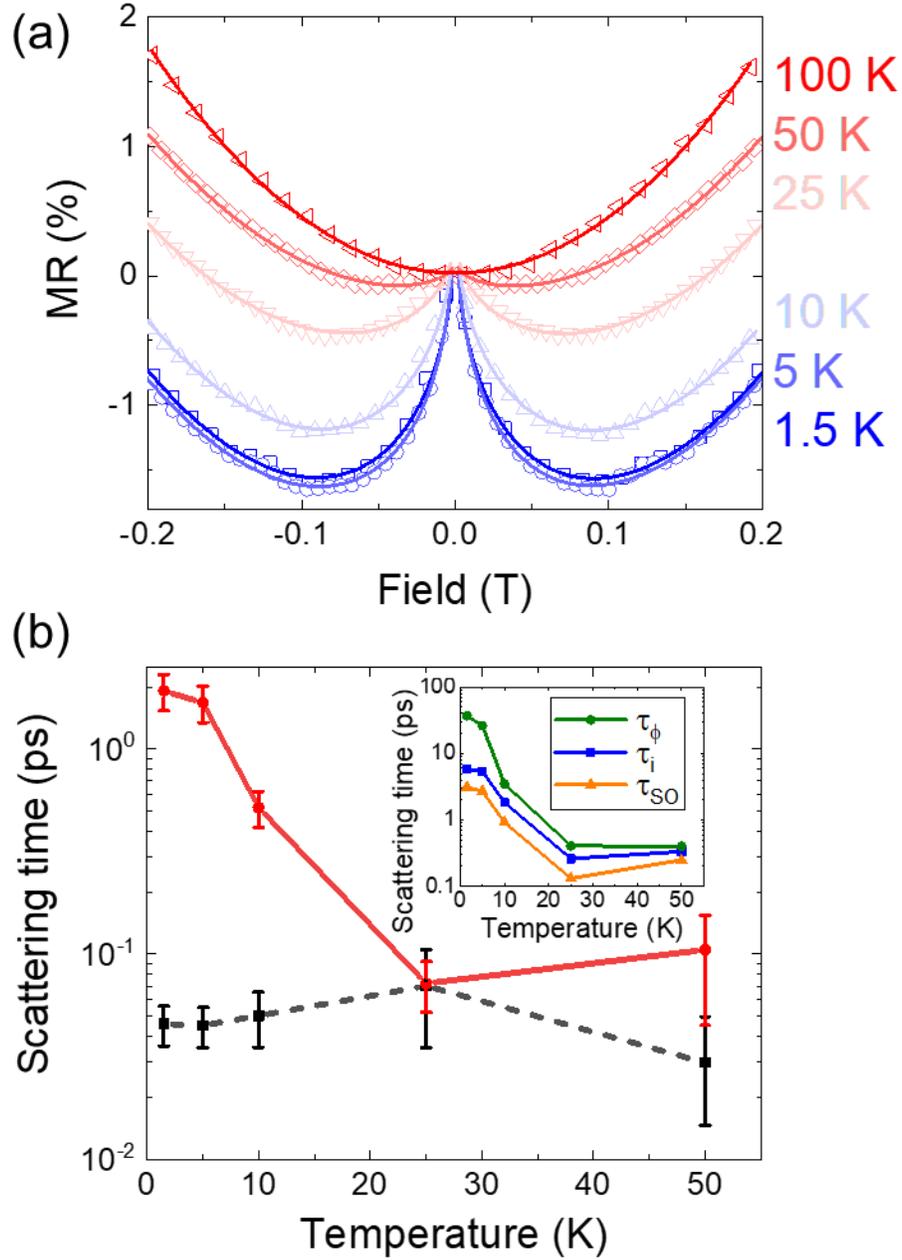

FIG. 4. (a) Temperature dependence of the WL peak in Gr/PST. The WL cusp is completely suppressed by 100 K. The data (symbols) are given along with the associated fits to Eq. 3 (lines). (b) Characteristic scattering times for WL ($\tau_{WL}$) and SdH ($\tau_{SdH}$) regimes as a function of temperature. There is a clear transition in the WL scattering time around 25 K, suggesting a topological phase transition. (inset) Scattering times obtained as parameters from WL fitting as a function of temperature. The scattering times in (inset) are added in reciprocal to get the characteristic WL scattering time in (b).

We now want to compare $\tau_{WL}$ with the high-field scattering time obtained from the SdH oscillations, $\tau_{SdH}$. Fig. 4(b) shows both $\tau_{SdH}$ and $\tau_{WL}$ at temperatures up to 50 K. Above, 25K, the peaks become difficult to distinguish from the background, increasing the uncertainty (See Supplemental Information Fig. S5). Overall, $\tau_{SdH}$ remains relatively constant around 700 fs, while $\tau_{WL}$ shows a distinct transition around 25 K. One would expect the SdH momentum scattering time to be an average of the scattering times obtained from the WL fits ($\tau_{SO} = 1000$ ps, $\tau_\phi = 7$ ps, $\tau_i = 13$ ps), as each mechanism in the WL model contributes to the momentum scattering events. This gives an estimate of the low-field scattering time of $\left(\tau_{SO}^{-1} + \tau_i^{-1} + \tau_\phi^{-1}\right)^{-1} = \tau_{WL} = 4.5$ ps, six times larger than the high field scattering time, $\tau_{SdH} = 0.72$ ps. This discrepancy in high- and low-field scattering times may be a result of spin splitting or field-induced asymmetry in the 2DEG bands. The magnetic field splits or skews the bands, altering the conduction paths and reducing the scattering time.

The 2DEG in Gr/PST is, however, analogous to the topologically protected surface states in topological insulators, like PST itself. In the case of topological materials, the surface states are protected by time reversal symmetry. Therefore, breaking time reversal symmetry, e.g. applying a magnetic field, destroys the topological state.[26] In the case of Gr/PST, the 2DEG is protected by a special symmetry relating to the polarization. The 2DEG could then be broken by defects or disorder that affects the polarization.

If we assume all scattering events are spin scattering events, the WL scattering time would be the spin scattering time. Therefore, we can use $\tau_{WL} = \tau_s$ as a lower bound on the spin lifetime. Assuming a D'yakonov-Perel (DP) spin scattering mechanism, as has been observed in topological materials like $Bi_2Se_3$ as well as graphene with induced SOC,[34,37] we can approximate the strength of the spin-orbit interaction in our Gr/PST heterostructure. For the DP mechanism, spin relaxation occurs due to precession in a random effective magnetic field. In that case, the momentum ($\tau_q$) and spin ($\tau_s$) scattering times are related by the spin-orbit interaction strength $\Delta_{SO}$ according to:

$$\tau_s^{DP} = \frac{\hbar^2}{4\Delta_{SO}^2 \tau_q}. \qquad (4)$$

This yields an effective spin-orbit interaction of $\Delta_{SO} = 0.3$ meV. This is significantly lower than the theoretical 33 meV interaction that can be induced in graphene by $WS_2$,[38] but on par with experimental results in $WS_2$/graphene spin valves,[34] and an order of magnitude larger than the 0.050 meV measured in suspended graphene devices.[15]

**Conclusions**

Magnetotransport provides an interesting picture of the interaction between graphene and PST, with strong evidence of a charge redistribution to avoid the polar catastrophe. This heterostructure system manifests as a high mobility 2D conduction channel with 50 times longer scattering times compared to graphene. These properties suggest the Gr/PST heterostructure is an ideal candidate for further investigation, particularly as a spin transport channel. It also suggests that the polar catastrophe model could precipitate 2DEGs in other similar heterostructures created through stacking dissimilar materials. Further study of this system, both by theory and through DOS-probing techniques such as ARPES, could provide further insight into interaction between topological and 2D materials. The improved mobility and scattering time suggest graphene/topological heterostructures could be utilized in a similar way as boron nitride encapsulation, but with additional properties advantageous to computational device applications.

**Methods/Experimental**

$Pb_{0.24}Sn_{0.76}Te$ (111) (PST) films were grown on GaAs (001) by MBE. Because the PST layers of interest are only 7 nm thick, we incorporated an electrically insulating chalcogenide-based buffer layer to decouple the interesting topological phenomenology of the PST from the unwanted effects of the GaAs substrate. Specifically, we used a sacrificial 10 nm $Bi_2Se_3$ starting layer which adopts a (001) trigonal growth orientation upon which we grew a 20 nm $(In,Bi)_2Se_3$ followed by a 20 nm $In_2Se_3$ layer whose large trivial energy gap (1.3 eV) and smooth surface morphology make it a preferred surface for exploring the transport properties of the PST. The (001) trigonal symmetry of the sacrificial $Bi_2Se_3$ seeds the (111) orientation for the final PST layer of interest. Upon heating to >500º C for an hour, all traces of $Bi_2Se_3$ completely evaporate, leaving the templated, atomically smooth $In_2Se_3$ upon which we use individual molecular beams of lead (Pb), tin (Sn), and tellurium (Te) to produce topological crystalline insulator (TCI) films having $Pb_{0.24}Sn_{0.76}Te$ composition. The stoichiometry is chosen to maximize the bulk band gap while maintaining the TCI behavior.[39] PST (111) bandstructure is strongly dependent on surface passivation.[33] Previous modeling showed two well-isolated Dirac cones, at Γ and M when the surface is passivated. However, for an unpassivated surface, the cones are not isolated and coexist with additional metallic energy bands. For our case, we assume that the bottom surface of the film is passivated by the substrate during MBE growth and the top surface is passivated due to sustained exposure to atmosphere. Likewise, the lack of observation of a strong electrically conducting state in bare PST corroborates these assumptions.

Films of PST were patterned into Hall bars using Shipley S1800 photoresist and optical lithography followed by etching in Ar/$H_2$ plasma. Another optical lithography step defined contacts, which were created with electron beam deposition of Ti/Au (5/50 nm) followed by lift-off in acetone. Graphene was grown by low-pressure (5-50 mTorr) chemical vapor deposition on copper foils at 1030°C under flowing $H_2$ and $CH_4$ gas.[40] After growth, the Cu foil substrates were etched and the remaining monolayer graphene films were transferred onto the predefined PST devices using a wet process (methods available elsewhere in the literature).[41] Another lithography step using polymethyl methacrylate (PMMA) and Deep-UV photolithography was performed to again define the Hall bar mesas. PMMA and Deep-UV lithography are used here to minimize chemical contamination of the graphene.[42] Excess graphene was subsequently removed with $O_2$ plasma.

The PST Hall bars were measured before and after the graphene transfer in order to directly compare changes in mobility and conductivity. Magnetotransport measurements were performed in a variable temperature $He^3$ refrigerator equipped with a sample rotator within a 12 T superconducting magnet. We utilized a lock-in amplifier with independent preamplifiers and filters to simultaneously measure the longitudinal and transverse magnetoresistance. Results were consistent across 5 devices made from two nominally identical PST growths and processed at different times.

Raman spectra were measured in the backscattering arrangement. The sample was placed in a closed-cycle variable-temperature (5-300 K) cryostat. We used a Kr/Ar ion laser (488 to 647 nm) and a 532 nm diode laser as the laser sources. The laser power was limited to < 10 kW/cm$^2$ with 30 s exposure time to maximize the peak intensity and minimize the damage to the sample. An 1800-grooves/mm grating and a Si CCD were used to diffract and detect the scattered light, respectively.

We performed density-functional theory (DFT) calculations for structures similar to Gr/PST heterostructures to further explain the experimental results. *Ab initio* calculations were carried out using the projector-augmented wave method[43] as implemented in the Vienna ab-initio simulation package (VASP).[44] See the Supplemental Information for further calculation details.

Due to the size of the supercell needed to accurately describe alloyed PST, calculations are limited to two extreme cases: (i) a heterostructure formed of graphene on purely topological SnTe, and (ii) heterostructures formed with the Sn-atoms in the top $N$ layers of SnTe replaced with Pb, denoted as $N$L-PbSnTe, *e.g.* 4L-PbSnTe has the top four layers of Sn replaced by Pb. The latter is equivalent to having $N$ layers of the trivial insulator PbTe on top of the TI SnTe. Our calculations (see Supplemental Information Fig. S2) support that our alloyed $Pb_{0.24}Sn_{0.76}Te$ has the Sn and Pb species evenly distributed, and lying somewhere between these two extremes, on the topological side of the phase diagram.[39] In both cases, we look at the change in electronic structure with and without ($\sqrt{3}$x$\sqrt{3}$) graphene on the (111) surface of the SnTe.

## Author Contributions

A.L.F. formulated the initial experimental idea. A.L.F. and G.M.S. designed the experiments with assistance from A.T.H. A.L.F fabricated the devices with assistance from P.M.C. J.T.R. grew and transferred graphene. P.J.T. grew the PST films. G.M.S. performed transport experiments and transport data analysis with assistance from J.E.D., N.A.B., P.M.C., A.T.H. and A.L.F. Raman spectroscopy, optical characterization, and analysis was performed by Y.-J.L.S. with assistance from J.T.R. and A.T.H. DFT modeling and analysis was performed by I.N., S.S. and P.D. G.M.S., A.T.H., I.N. and A.L.F. wrote the manuscript. All authors contributed to discussions of results and editing the manuscript.

## Supplementary Information

As-measured Raman spectra, further calculations on Pb/Sn alloying, projected DoS for the bulk and surface atomic species, information on Shubnikov de Haas background subtraction, as well as fits for the effective mass and scattering times can be found in the Supplementary materials available online.

## Acknowledgements

The authors from LPS gratefully acknowledge critical assistance from LPS support staff including G. Latini, J. Wood, R. Brun, P. Davis, and D. Crouse. The authors from LPS and ARL along with PMC gratefully acknowledge support from the Applied Research for the Advancement of S&T

Priorities program of the Office of the Secretary of Defense. Work at NRL was supported by the Office of Naval Research through Base programs at NRL. PD, IN, and SS acknowledge support by the W. M. Keck Foundation and the NSF Grant number DMR-1752840. The computational resources were provided by the Extreme Science and Engineering Discovery Environment (XSEDE) under Project PHY180014, which was supported by National Science Foundation grant number ACI-1548562.

**Competing financial interests**

The authors declare no competing financial interests.